\documentclass[prb,twocolumn,showpacs,amsmath,amssymb]{revtex4}

\usepackage{graphicx}
\usepackage{dcolumn}
\usepackage{bm}

\begin{document}
\title
{
Entropy and Spin Susceptibility of $s$-wave Type-II Superconductors near $H_{c2}$
}

\author{Takafumi Kita}
\affiliation{Division of Physics, Hokkaido University, Sapporo 060-0810,
Japan}

\date{\today}

\begin{abstract}
A theoretical study is performed on the entropy $S_{\rm s}$ and 
the spin susceptibility $\chi_{\rm s}$
near the upper critical field $H_{c2}$
of $s$-wave type-II superconductors with arbitrary impurity concentrations.
The changes of these quantities through $H_{c2}$ may be expressed as 
$[S_{\rm s}(T,B)\!-\!S_{\rm s}(T,0)]/[S_{\rm n}(T)\!-\!S_{\rm s}(T,0)]\!=\!1\!-\!\alpha_{S}(1\!-\!B/H_{c2})\!\approx\!(B/H_{c2})^{\alpha_{S}}$,
for example, 
where $B$ is the average flux density and $S_{\rm n}$ denotes entropy in the normal state.
It is found that the slopes $\alpha_{S}$ and $\alpha_{\chi}$ at $T\!=\!0$ are identical,
connected directly with the zero-energy density of states, and vary from
$1.72$ in the dirty limit to $0.5\!\sim\!0.6$ in the clean limit.
This mean-free-path dependence of $\alpha_{S}$ and $\alpha_{\chi}$ at $T\!=\!0$ is
quantitatively the same as that of the slope $\alpha_{\rho}(T\!=\!0)$ for 
the flux-flow resistivity studied previously.
The result suggests that $S_{\rm s}(B)$ and 
$\chi_{\rm s}(B)$ near $T\!=\!0$ are
convex downward (upward) in the dirty (clean) limit,
deviating substantially from the linear behavior $\propto\!B/H_{c2}$.
The specific-heat jump at $H_{c2}$ also shows fairly large mean-free-path dependence.
\end{abstract}
\maketitle

\section{Introduction}
\label{sec:intro}

This paper considers the
changes of the entropy $S_{\rm s}$ and 
the spin susceptibility $\chi_{\rm s}$ through
$H_{c2}$ for classic $s$-wave type-II superconductors.
These quantities were calculated by Maki\cite{Maki65,Maki66} 
in the dirty limit for superconducting alloys nearly 40 years ago.
However, detailed studies on clean systems are still missing
even for $s$-wave superconductors.
Writing these quantities as
\begin{subequations}
\label{alpha}
\begin{equation}
\frac{S_{\rm s}(T,B)\!-\!S_{\rm s}(T,0)}{S_{\rm n}(T)\!-\!S_{\rm s}(T,0)}=
1\!-\alpha_{S}\!\left(\!1\!-\!\frac{B}{H_{c2}}\!\right)\approx
\left(\!\frac{B}{H_{c2}}\!\right)^{\!\!\alpha_{S}}\!\! ,
\label{alphaS}
\end{equation}
\begin{equation}
\frac{\chi_{\rm s}(T,B)\!-\!\chi_{\rm s}(T,0)}{\chi_{\rm n}(T)\!-\!\chi_{\rm s}(T,0)}=
1\!-\alpha_{\chi}\!\left(\!1\!-\!\frac{B}{H_{c2}}\!\right)\approx
\left(\!\frac{B}{H_{c2}}\!\right)^{\!\!\alpha_{\chi}}\!\! ,
\label{alphaC}
\end{equation}
\end{subequations}
the slopes $\alpha_{S}$ and $\alpha_{\chi}$ will be obtained quantitatively
for arbitrary impurity concentrations.
The results near $H_{c2}$ will also be useful for getting an insight into the behaviors
over $0\!\leq\! B\!\leq\! H_{c2}$.
Indeed, $\alpha\!>\! 1$ ($\alpha\!<\! 1$) indicates overall field dependence 
which is convex downward (upward), as realized from Eq.\ (\ref{alpha}).

It seems to have been widely accepted
that various physical quantities of classic $s$-wave type-II superconductors
follow the linear field dependence with $\alpha\!=\!1$
at low temperatures.
A theoretical basis for it is
the density of states for a single vortex calculated
by Caroli, de Gennes, and Matricon.\cite{CdGM64,Fetter69}
However, few quantitative calculations have been carried out so far 
on the explicit field dependence.
Recently, Ichioka {\em et al}.\cite{Ichioka99} performed a numerical study on the density of
states of clean two-dimensional $s$-wave superconductors 
with $\kappa\!\gg\! 1$ at $T\!=\!0.5T_{c}$.
They found the exponent $\alpha\!=\!0.67$ for the overall field dependence 
of the zero-energy density of states.
Also, experiments on the $T$-linear specific-heat coefficient $\gamma_{\rm s}(B)$ for clean 
V$_{3}$Si,\cite{Ramirez96}
NbSe$_{2}$,\cite{Sanchez95,Nohara99,Sonier99,Hanaguri03} and CeRu$_{2}$\cite{Hedo98}
show marked upward deviations
from the linear behavior $\gamma_{n}B/H_{c2}$.
Even early experiments on $\gamma_{\rm s}(B)$ for clean
V and Nb indicate similar deviations,\cite{Radebaugh66,Ferreira69}
although not recognized explicitly in those days
due to the absence of a theory on clean systems.
These results indicate that the field dependence with $\alpha\!<\! 1$
is a general feature of clean $s$-wave superconductors, as suggested
by Ramirez.\cite{Ramirez96}

Following the preceding works on the Maki para\-meters\cite{Kita03}
and the flux-flow resistivity,\cite{Kita04}
which will be referred to as I and II, respectively,
I here present a detailed study on $S_{\rm s}$ and $\chi_{\rm s}$
near $H_{c2}$ at all temperatures.
I thereby hope to clarify the $\kappa$ and mean-free-path ($l_{\rm tr}$) dependence 
of $\alpha_{S}$ and $\alpha_{\chi}$.
Calculations are performed for both two and three dimensional isotropic systems 
to see the dependence of $\alpha_{S}$ and $\alpha_{\chi}$
on detailed Fermi-surface structures.
I also calculate the specific heat jump at $H_{c2}$
for various values of $\kappa$ and $l_{\rm tr}$.
To my knowledge, this kind of a systematic study has not been performed
even for classic $s$-wave superconductors.

Unlike the convention, I adopt the average flux density $B$ in the bulk
as an independent variable instead of the external field $H$.
An advantage for it is that the irrelevant region $H\!\leq\!H_{c1}$ is automatically removed
from the discussion on the field dependence.
This distinction between $B$ and $H$ becomes important for low-$\kappa$ materials
where $H\!\leq\!H_{c1}$ occupies a substantial part
of $0\!\leq\! H\!\leq\! H_{c2}$. 
Any experiment on the $B$ dependence should be accompanied 
by a careful measurement on the magnetization,
especially for low-$\kappa$ materials
like Nb and V.

Section \ref{sec:formulation} provides the
formulation,
Sec \ref{sec:results} presents numerical results,
and 
Sec.\ \ref{sec:summary} summarizes the paper.
I put $k_{\rm B}\!=\!1$ throughout.

\section{Formulation}
\label{sec:formulation}
\subsection{Entropy and Pauli paramagnetism}

As before,\cite{Kita03,Kita04} I consider the $s$-wave pairing 
with an isotropic Fermi surface and $s$-wave 
impurity scattering in an external magnetic field ${\bf H}\parallel \!{\bf z}$.
The formulation proceeds in exactly the same way for both 
the three dimensional system and the two-dimensional system placed
in the $xy$ plane perpendicular to ${\bf H}$.
The vector potential in the bulk can be written 
as\cite{Eilenberger64,Lasher,Marcus,Brandt,DGR90,Kita98}
\begin{equation}
{\bf A}({\bf r})\!=\! Bx\hat{\bf y}\!+\tilde{\bf A}({\bf r})\, ,
\label{A}
\end{equation}
where $B$ is the average flux density produced jointly by the external current 
and the supercurrent inside the sample, and
$\tilde{\bf A}$ expresses the spatially varying part of the magnetic field 
satisfying $\int{\bm \nabla}\!\times\!\tilde{\bf A}\,d{\bf r}\!=\!{\bf 0}$.

I first write down the expressions of the entropy and the magnetization in the presence
of Pauli paramagnetism.
As can be checked directly,\cite{SR83} the effect can be included 
in the Eilenberger equations\cite{Eilenberger68} for the quasiclassical Green's
functions $f$, $f^{\dagger}$, and $g$ by the replacement:
\begin{equation}
\varepsilon_{n} \rightarrow \varepsilon_{n}'\equiv \varepsilon_{n}-i\mu_{\rm B}
\,\hat{\bf z}\!\cdot\!({\bm \nabla}\!\times\!{\bf A})\, ,
\label{varepsilon_n'}
\end{equation}
where $\varepsilon_{n}\!\equiv\!(2n\!+\!1)\pi T$ is the Matsubara energy and
$\mu_{\rm B}$ is the Bohr magneton.
The corresponding Eilenberger functional\cite{Eilenberger68} 
for the free-energy difference between the
normal and superconducting states is given by
\begin{eqnarray}
&&\hspace{-7mm}F
=\int\!d{\bf r}
\biggl\{ \frac{
({\bm \nabla}\!\times\!{\bf A})^{2}}{8\pi}+N(0)|\Delta({\bf r})|^{2}\ln\frac{T}{T_{c}}
\nonumber \\
&&+\pi T N(0) \sum_{n=-\infty}^{\infty}\left[\frac{|\Delta({\bf r})|^{2}}{|\varepsilon_{n}|}-
\langle I(\varepsilon_{n},{\bf k}_{\rm F},{\bf r})\rangle \right]
\biggr\} .
\label{F}
\end{eqnarray}
Here $\Delta$ is the pair potential, 
$N(0)$ is the density of states per spin and per unit volume
at the Fermi level, 
${\bf k}_{\rm F}$ is the Fermi wavevector,
and $\langle \cdots \rangle$ denotes
Fermi-surface average satisfying $\langle 1 \rangle\!=\! 1$.
The quantity $I$ is defined by\cite{Kita03}
\begin{eqnarray}
&&\hspace{-7mm}I\equiv \Delta^{\! *}f\!+\!\Delta f^{\dagger} 
+2\varepsilon_{n}'[g\!-\!{\rm sgn}
(\varepsilon_{n})]
+\hbar\,\frac{f\langle f^{\dagger}\rangle\!+\!\langle f\rangle f^{\dagger}}{4\tau}
\nonumber \\
&&\hspace{-1mm}
+\hbar\,\frac{g\langle g\rangle\!-\! 1}{2\tau}
-\hbar\,\frac{f^{\dagger}\,{\bf v}_{\rm F}\!\cdot\!
{\bm \partial}f
-f\,{\bf v}_{\rm F}\!\cdot\!
{\bm \partial}^{*}f^{\dagger}}
{2[g\!+\!{\rm sgn}
(\varepsilon_{n})]} \, ,
\label{I}
\end{eqnarray}
where $\tau$ is the relaxation time in the second-Born approximation,
${\bf v}_{\rm F}$ is the Fermi velocity, and ${\bm \partial}$ denotes
\begin{equation}
{\bm\partial}\equiv{\bm \nabla}-i\frac{2e}{\hbar c}{\bf A}\, .
\end{equation}
The quasiclassical Green's functions $f$ and $g$ are connected by
$g\!=\!(1\!-\! ff^{\dagger})^{1/2}{\rm sgn}(\varepsilon_{n})$
with $f^{\dagger}(\varepsilon_{n},{\bf k}_{\rm F},{\bf r};\mu_{\rm B})\!=
\!f^{*}(\varepsilon_{n},-{\bf k}_{\rm F},{\bf r};-\mu_{\rm B})$.
The change of sign in $\mu_{\rm B}$ is necessary here,
because $f\!\equiv\!f_{\uparrow\downarrow}\!\neq
\!f_{\downarrow\uparrow}$ in the presence of Pauli paramagnetism.
The functional derivatives of Eq.\ (\ref{F}) with respect to $f^{\dagger}$, 
$\Delta^{\!*}$, and $\tilde{\bf A}$ lead to the Eilenberger equation for $f$,
the self-consistency equation for $\Delta({\bf r})$,
and the Maxwell equation for $\tilde{\bf A}$, respectively.

The expression of the entropy $S_{\rm s}$
is obtained from Eq.\ (\ref{F}) by the thermodynamic relation:
$S_{\rm s}\!=\!S_{\rm n}\!- \partial F/\partial T$.
Considering the stationarity with respect to $f$, $\Delta$, and $\tilde{\bf A}$, 
we only have to differentiate with respect to the
explicit temperature dependence in $F$. 
We thereby obtain
\begin{eqnarray}
&&\hspace{-4mm}S_{\rm s}=S_{\rm n}-\frac{N(0)}{T}\int d{\bf r} \biggl[|\Delta({\bf r})|^{2}
-\pi T\sum_{n=-\infty}^{\infty}
\langle I(\varepsilon_{n},{\bf k}_{\rm F},{\bf r})\rangle
\nonumber \\
&&\hspace{20mm}
-2\pi T\sum_{n=-\infty}^{\infty}\varepsilon_{n}\langle g\!-\!{\rm sgn}
(\varepsilon_{n})\rangle\biggr] \, ,
\label{S}
\end{eqnarray}
where $S_{\rm n}\!=\!2\pi^{2}N(0)VT/3$ with 
$V$ the volume of the system.
In contrast, the expression of the external field $H$ may be derived 
by applying the Doria-Gubernatis-Rainer scaling to Eq.\
(\ref{F}).\cite{DGR89}
The details are given in Appendix A of I,
and we obtain
\begin{eqnarray}
&&\hspace{-4mm}H=-4\pi M_{{\rm nP}}+
B+\frac{1}{BV}\int\! d{\bf r}\,({\bm \nabla}\!\times\!\tilde{\bf A})^{2}
\nonumber \\
&&\hspace{-4mm}+\frac{\pi^{2} TN(0)}{BV}\sum_{n=-\infty}^{\infty}\int \! d{\bf r} \left<
\hbar\,\frac{f^{\dagger}\,{\bf v}_{\rm F}\!\cdot\!
{\bm \partial}f
-f\,{\bf v}_{\rm F}\!\cdot\!
{\bm \partial}^{*}f^{\dagger}}
{g\!+\!{\rm sgn}
(\varepsilon_{n})}\right>
\nonumber \\
&&\hspace{-4mm}+i\frac{8\pi^{2} TN(0)\mu_{\rm B}}{BV}
\sum_{n=-\infty}^{\infty}\int \! d{\bf r}\,
\langle g\rangle \,
\hat{\bf z}\!\cdot\!({\bm \nabla}\!\times\!{\bf A})
\, ,
\label{H-B}
\end{eqnarray}
where $M_{{\rm nP}}\!=\!2\mu_{\rm B}^{2}N(0)B$ denotes 
the normal-state magnetization due to Pauli paramagnetism.
We thus arrives at the expression of the magnetization from Pauli paramagnetism as
\begin{equation}
M_{{\rm sP}}=M_{{\rm nP}}-i\frac{2\pi TN(0)\mu_{\rm B}}{BV}\!\!
\sum_{n=-\infty}^{\infty}\!\int \! d{\bf r}\,
\langle g\rangle \, \hat{\bf z}\cdot({\bm \nabla}\!\times\!{\bf A})\, .
\label{Ms}
\end{equation}

When Pauli paramagnetism is negligible compared with the
orbital diamagnetism by supercurrent,
we can take the limit $\mu_{B}\!\rightarrow\! 0$ in Eqs.\ (\ref{S}) and (\ref{Ms})
and retain only the leading-order terms.
This results in $\varepsilon_{n}'\!\rightarrow\!\varepsilon_{n}$ for Eq.\ (\ref{S}).
On the other hand, Eq.\ (\ref{Ms}) is transformed by noting Eq.\ (\ref{varepsilon_n'}) into
\begin{equation}
M_{{\rm sP}}=M_{{\rm nP}}\left[1-\frac{\pi T}{V}\!\!
\sum_{n=-\infty}^{\infty}\!\int \! d{\bf r}\,
\frac{\partial  \langle g\rangle}{\partial \varepsilon_{n}}\!
\left(\!\frac{{\bm \nabla}\!\times\!{\bf A}}{B}\!\right)^{\!\!2}\,
\right] \, .
\label{Ms1}
\end{equation}
If the zero-field expression 
$g\!=\!\varepsilon_{n}/\sqrt{\varepsilon_{n}^{2}+|\Delta|^{2}}$
is substituted into Eq.\ (\ref{Ms1})
with ${\bm \nabla}\!\times\!{\bf A}\!=\!B\hat{\bf z}$, the terms in the square bracket
reduces to the Yosida function.\cite{Yosida58}

\subsection{Expressions near $H_{c2}$}

I now consider the cases where Pauli paramagnetism is small
and provide explicit expressions to Eqs.\ (\ref{S}) and (\ref{Ms1}) near $H_{c2}$.
From now on I adopt the units used previously\cite{Kita03,Kita04}
where the energy, the length, and the magnetic field
are measured by the zero-temperature energy gap $\Delta(0)$ at $H\!=\!0$,
the coherence length $\xi_{0}\!\equiv\!\hbar v_{\rm F}/\Delta(0)$ 
with $v_{\rm F}$ the Fermi velocity, 
and $B_{0}\!\equiv\!\phi_{0}/2\pi\xi_{0}^{2}$ with 
$\phi_{0}\!\equiv\! hc/2e$ the flux quantum,
respectively, with $\hbar\!=\! 1$.

First, $f$, $g$, and $\tilde{\bf A}$ are expanded up 
to the second order in $\Delta({\bf r})$ as
\begin{equation}
\left\{
\begin{array}{l}
\vspace{2mm}
f=f^{(1)}
\\
\vspace{2mm}
g= \left(\,1- \frac{1}{2}f^{(1)\dagger}f^{(1)}\,\right)\,{\rm sgn}(\varepsilon_{n})
\\ 
\tilde{\bf A}\!=\!\tilde{\bf A}^{(2)}
\end{array}
\right. .
\label{fgAExp}
\end{equation}
Substituting them into Eqs.\ (\ref{S}) and (\ref{Ms1}) and using the Eilenberger equations
for $f^{(1)}$ and $f^{(1)\dagger}$ to remove terms with ${\bf v}_{\rm F}\!\cdot\!{\bm \partial}$, 
we obtain
\begin{subequations}
\label{S-M2}
\begin{eqnarray}
&&\hspace{-8mm}
\frac{S_{\rm s}}{S_{\rm n}}=1-\frac{3}{2\pi^{2}T^{2}V}\!\int d{\bf r} \biggl[ |\Delta({\bf r})|^{2}\!
-\frac{\pi T}{2}\sum_{n}
\langle f^{(1)\dagger}\Delta
\nonumber \\
&&\hspace{10mm}
+f^{(1)}\Delta^{*} \rangle+\pi T\sum_{n}|\varepsilon_{n}|\,\langle f^{(1)\dagger}f^{(1)}\rangle
\biggr] \, ,
\label{S2}
\end{eqnarray}
\begin{equation}
\frac{M_{{\rm sP}}}{M_{{\rm nP}}}=1+\frac{\pi T}{2V}\!
\sum_{n}\!\int \! d{\bf r}\,
\biggl<\!\frac{\partial  f^{(1)\dagger}}{\partial \varepsilon_{n}}f^{(1)}\!+\!
f^{(1)\dagger}\,\frac{\partial  f^{(1)}}{\partial \varepsilon_{n}}\!\biggr> 
{\rm sgn}(\varepsilon_{n}) \, .
\label{Ms2}
\end{equation}
\end{subequations}
Further, $\Delta({\bf r})$ and $f^{(1)}$ near $H_{c2}$
can be expanded in the basis functions $\psi_{N{\bf q}}({\bf r})$ of 
the vortex lattice as\cite{Kita03}
\begin{subequations}
\label{Expand}
\begin{equation}
\Delta({\bf r})
=\sqrt{V}\Delta_{0}\,\psi_{0{\bf q}}({\bf r}) \, ,
\label{DeltaExpand}
\end{equation}
\begin{equation}
f^{(1)}(\varepsilon_{n},{\bf k}_{\rm F},{\bf r})
=\sqrt{V}\Delta_{0}
\sum_{N=0}^{\infty} \tilde{f}_{N}^{(1)}(\varepsilon_{n},\theta)\,
{\rm e}^{i N\varphi}\,\psi_{N{\bf q}}({\bf r}) 
\, ,
\label{fExpand}
\end{equation}
\end{subequations}
where $(\theta,\varphi)$ are the polar angles of ${\bf v}_{\rm F}$
with $\sin\theta\!\rightarrow\! 1$ in two dimensions,
$N$ denotes the Landau level, and
${\bf q}$ is an arbitrary chosen magnetic Bloch vector 
characterizing the broken translational symmetry of the flux-line lattice
and specifying the core locations.\cite{Kita98}
The coefficients $\Delta_{0}$ and $\tilde{f}_{N}^{(1)}$ are both real for the 
relevant hexagonal lattice.
Substituting these expressions into Eqs.\ (\ref{S2}) and (\ref{Ms2})
and using the orthonormality of $\psi_{N{\bf q}}({\bf r})$ and ${\rm e}^{i N\varphi}$,
we obtain
\begin{subequations}
\label{S-M3}
\begin{eqnarray}
&&\hspace{-10mm}
\frac{S_{\rm s}}{S_{\rm n}}=1-\frac{3\Delta_{0}^{2}}{2\pi^{2}T^{2}} \biggl[ 1
-\pi T\sum_{n=-\infty}^{\infty}
\langle \tilde{f}^{(1)}_{0} \rangle
\nonumber \\
&&\hspace{5mm}
+\pi T\sum_{n=-\infty}^{\infty}|\varepsilon_{n}|\,
\sum_{N}\,(-1)^{N}
\langle \tilde{f}^{(1)}_{N}\tilde{f}^{(1)}_{N}\rangle \biggr] \, ,
\label{S3}
\end{eqnarray}
\begin{equation}
\frac{M_{{\rm sP}}}{M_{{\rm nP}}}=1+\pi T\Delta_{0}^{2}
\sum_{n=-\infty}^{\infty}\sum_{N}\,(-1)^{N}
\biggl<\frac{\partial  \tilde{f}^{(1)}_{N}}{\partial \varepsilon_{n}}\tilde{f}^{(1)}_{N}
\!\biggr> 
\, {\rm sgn}(\varepsilon_{n}) \, .
\label{Ms3}
\end{equation}
\end{subequations}
Except $\Delta_{0}^{2}\!\propto\!H_{c2}\!-\!B$,
all the quantities in Eqs.\ (\ref{S3}) and (\ref{Ms3}) are to be evaluated at $H_{c2}$.

It is possible to give an alternative expression
to Eq.\ (\ref{S3}) using
the equation for $H_{c2}$ given by Eq.\ (33) of I:
\begin{equation}
\ln \frac{T_{c}}{T}+\pi T \sum_{n=-\infty}^{\infty}
\left[\langle 
\tilde{f}_{0}^{(1)}(\varepsilon_{n})\rangle-\frac{1}{|\varepsilon_{n}|}\right] =0 \, .
\label{Hc2}
\end{equation}
Differentiating Eq.\ (\ref{Hc2}) with respect to $T$ yields
\begin{equation}
-1+\pi T \sum_{n}\left[\langle 
\tilde{f}_{0}^{(1)}\rangle+
\frac{\partial\langle 
\tilde{f}_{0}^{(1)}\rangle}{\partial\varepsilon_{n}}\varepsilon_{n}
+\frac{\partial\langle 
\tilde{f}_{0}^{(1)}\rangle}{\partial H_{c2}}T\frac{d H_{c2}}{d T}
\right] \! = 0 \, .
\label{dHc2}
\end{equation}
The quantity ${\partial\langle 
\tilde{f}_{0}^{(1)}\rangle}/{\partial H_{c2}}$ has been calculated as Eqs.\ (31)-(32) 
of I to be
\begin{equation}
\frac{\partial\langle 
\tilde{f}_{0}^{(1)}\rangle}{\partial H_{c2}} =\sum_{N}
(-1)^{N+1}\sqrt{\frac{N+1}{8H_{c2}}}\,
\langle\tilde{f}^{(1)}_{N+1}\tilde{f}^{(1)}_{N}\sin\theta 
\rangle \, . 
\label{df_0/dB}
\end{equation}
A similar procedure enables us to obtain the
expressions of ${\partial\langle 
\tilde{f}_{0}^{(1)}\rangle}/{\partial \varepsilon_{n}}$ and ${\partial
\tilde{f}_{N}^{(1)}}/{\partial \varepsilon_{n}}$ in Eq.\ (\ref{Ms3}) as
\begin{subequations}
\label{f_0N'}
\begin{equation}
\frac{\partial\langle 
\tilde{f}_{0}^{(1)}\rangle}{\partial \varepsilon_{n}} = -\sum_{N}\,(-1)^{N}
\langle \tilde{f}^{(1)}_{N}\tilde{f}^{(1)}_{N}\rangle\, {\rm sgn}(\varepsilon_{n}) \, ,
\label{f_0'}
\end{equation}
\begin{equation}
\frac{\partial
\tilde{f}_{N}^{(1)}}{\partial \varepsilon_{n}} = -\sum_{N}\,
K^{N'}_{N}\tilde{f}^{(1)}_{N'}+\frac{K^{0}_{N}}{2\tau}{\rm sgn}(\varepsilon_{n})
\frac{\partial\langle 
\tilde{f}_{0}^{(1)}\rangle}{\partial \varepsilon_{n}} \, ,
\label{f_N'}
\end{equation}
\end{subequations}
where $K^{N'}_{N}$ is defined by Eq.\ (25) of I.
Using Eqs.\ (\ref{dHc2}) and (\ref{f_0'}) in Eq.\ (\ref{S3}), we obtain
\begin{equation}
\frac{S_{\rm s}}{S_{\rm n}}=1-\frac{d H_{c2}}{d T} \, \frac{3\Delta_{0}^{2}}{2\pi} 
\sum_{n=-\infty}^{\infty} \frac{\partial\langle 
\tilde{f}_{0}^{(1)}(\varepsilon_{n})\rangle}{\partial H_{c2}} \, ,
\label{S3b}
\end{equation}
with
\begin{equation}
\frac{d H_{c2}}{d T}=\frac{\displaystyle
1-\pi T \sum_{n=-\infty}^{\infty}\left[\langle 
\tilde{f}_{0}^{(1)}\rangle+
\frac{\partial\langle 
\tilde{f}_{0}^{(1)}\rangle}{\partial\varepsilon_{n}}\varepsilon_{n}\right]}
{\displaystyle
\pi T^{2} \sum_{n=-\infty}^{\infty}\frac{\partial\langle 
\tilde{f}_{0}^{(1)}\rangle}{\partial H_{c2}}}
\, .
\label{dHc2dT}
\end{equation}
Using Eq.\ (\ref{dHc2dT}) we can also calculate the specific-heat jump at
$H_{c2}$. It is given in conventional units as\cite{Maki65}
\begin{equation}
\Delta C = \frac{T}{4\pi}\!\left(\!\frac{dH_{c2}}{dT}\!\right)^{\!\! 2}
\frac{1}{(2\kappa_{2}^{2}\!-\!1)\beta_{\rm A}} \, ,
\label{deltaC}
\end{equation}
where $\kappa_{2}$ is the Maki parameter\cite{Maki64,Kita03} and $\beta_{\rm A}\!=\! 1.16$.

Equations (\ref{S-M3}) and (\ref{deltaC})
with Eqs.\ (\ref{df_0/dB}), (\ref{f_0N'}), 
and (\ref{dHc2dT}) are the main analytic results of the paper.
The quantities $\Delta_{0}$, $\tilde{f}^{(1)}_{N}$, and $\kappa_{2}$
have been obtained in I.
The explicit expression of $\tilde{f}^{(1)}_{N}$ is given by
\begin{equation}
\tilde{f}^{(1)}_{N}=\frac{{\tilde K}^{0}_{N}{\rm sgn}(\varepsilon_{n})}
{1-\langle {\tilde K}^{0}_{0}\rangle
{\rm sgn}(\varepsilon_{n})/2\tau} \, ,
\label{f_N}
\end{equation}
where ${\tilde K}^{N'}_{N}$ may be calculated efficiently by the procedure in Sec.\ IIF of 
I,
with a change of definition of $\tilde{\varepsilon}_{n}$ as
\begin{equation}
\tilde{\varepsilon}_{n}\equiv \left(\!|{\varepsilon}_{n}|
+\frac{1}{2\tau}\!\right){\rm sgn}(\varepsilon_{n}) \, .
\label{tilde-varepsilon}
\end{equation}

\subsection{Analytic results at $T=0$}

Now it will be shown that Eqs.\ (\ref{S3}) and (\ref{Ms3}) 
reduce to an identical expression at $T\!=\!0$
for arbitrary impurity concentrations, 
which has the physical meaning of the zero-energy density of states.

Let us start from Eq.\ (\ref{S3}) where 
$\varepsilon_{n}\!>\!0$ and $\varepsilon_{n}\!<\!0$ yield the same contribution.
Using this fact and Eq.\ (\ref{f_0'}),
it is transformed into
\begin{equation}
\frac{S_{\rm s}}{S_{\rm n}}=1-\frac{3\Delta_{0}^{2}}{2\pi^{2}T^{2}} \left[ 1
-2\pi T\sum_{n=0}^{\infty}\left(\!
\langle \tilde{f}^{(1)}_{0} \rangle+\varepsilon_{n}
\frac{\partial  \langle \tilde{f}^{(1)}_{0}\rangle}
{\partial \varepsilon_{n}}\!\right)\right] .
\label{S3a}
\end{equation}
The summation over $n$ for $T\!\rightarrow\!0$ may be performed 
by using the Euler-Maclaurin formula and 
the asymptotic property
$\tilde{f}^{(1)}_{0}(\varepsilon_{n})\!\rightarrow\!\varepsilon_{n}^{-1}$ 
($\varepsilon_{n}\!\rightarrow\!\infty$).\cite{Kita03}
For example,
\begin{eqnarray}
&&\hspace{-3mm}
2\pi T\sum_{n=0}^{\infty}
\langle \tilde{f}^{(1)}_{0}(\varepsilon_{n}) \rangle 
\nonumber \\
&&\hspace{-3mm}
\approx\int_{\pi T}^{\infty}\langle \tilde{f}^{(1)}_{0}(\varepsilon) \rangle \,d\varepsilon
+\pi T\langle \tilde{f}^{(1)}_{0}(\pi T) \rangle-\frac{(\pi T)^{2}}{3}
\langle \tilde{f}^{(1)\prime}_{0}(\pi T) \rangle
\nonumber \\
&&\hspace{-3mm}
\approx\int_{0}^{\infty}\langle \tilde{f}^{(1)}_{0}(\varepsilon) \rangle \,d\varepsilon
+\frac{(\pi T)^{2}}{6}
\langle \tilde{f}^{(1)\prime}_{0}(0) \rangle \, .
\end{eqnarray}
We thereby obtain 
\begin{subequations}
\label{S-M4}
\begin{equation}
\frac{S_{\rm s}}{S_{\rm n}}\,\stackrel{T\rightarrow 0}{\longrightarrow}\,
1+\frac{\Delta_{0}^{2}}{2} \langle \tilde{f}^{(1)\prime}_{0}(0)\rangle \, .
\label{S4}
\end{equation}
Equation (\ref{Ms3}) may be transformed similarly as
\begin{eqnarray}
&&\hspace{-3mm}\frac{M_{{\rm sP}}}{M_{{\rm nP}}}=1-\pi T\Delta_{0}^{2}
\sum_{n=0}^{\infty}
\frac{\partial^{2}  \langle \tilde{f}^{(1)}_{0}(\varepsilon_{n})\rangle}
{\partial \varepsilon_{n}^{2}}
\nonumber \\
&&\hspace{2.5mm}\stackrel{T\rightarrow 0}{\longrightarrow} 
1+\frac{\Delta_{0}^{2}}{2} \langle \tilde{f}^{(1)\prime}_{0}(0)\rangle
 \, .
\label{Ms4}
\end{eqnarray}
\end{subequations}
Thus, ${S_{\rm s}}/{S_{\rm n}}\!=\!{M_{\rm sP}}/{M_{\rm nP}}$, or equivalently, 
$\alpha_{S}\!=\!\alpha_{\chi}$ at $T\!=\!0$ for arbitrary impurity concentrations.

Equations (\ref{S4}) and (\ref{Ms4}) have a simple physical meaning.
Indeed, noting Eq.\ (\ref{fgAExp}), (\ref{fExpand}), and Eq.\ (\ref{f_0'}), 
we find an alternative expression at $T\!=\! 0$:
\begin{equation}
\frac{S_{\rm s}}{S_{\rm n}} = \frac{M_{\rm sP}}{M_{\rm nP}}=\frac{1}{V}\!\int
\langle g(\varepsilon_{n}\!=\!0,{\bf k}_{\rm F},{\bf r})\rangle \,  d{\bf r}
\, ,
\label{MS-DOS}
\end{equation}
which is nothing but the normalized density of states at 
$\varepsilon\!=\! 0$.
Thus, we have arrived at a simple result that 
the entropy and spin susceptibility at $T\!=\!0$ are both determined by
the zero-energy density of states.

The coefficient of $\Delta_{0}^{2}\!\propto\!H_{c2}\!-\!B$ have been obtained in I.
Also, $\tilde{f}^{(1)\prime}_{0}(0)$ in Eqs.\ (\ref{S4}) and (\ref{Ms4})
may be calculated efficiently from Eq.\ (\ref{f_N}) 
by using the analytic expression:\cite{Kita03}
\begin{eqnarray}
\tilde{K}^{0}_{0}(\tilde{\varepsilon}_{n},\beta) = 
\sqrt{\frac{2}{\pi}}\int_{0}^{\infty}\!\frac{\tilde{\varepsilon}_{n}}{
\tilde{\varepsilon}_{n}^{2}+x^{2}\beta^{2}}\,{\rm e}^{-x^{2}/2}\,dx  \, ,
\label{K00-2}
\end{eqnarray}
with $\beta\!\equiv\! {\sqrt{H_{c2}}\sin\theta}/{2\sqrt{2}}$.
Hence, Eqs.\ (\ref{S4}) and (\ref{Ms4}) at $T\!=\!0$ can be evaluated
fairly easily.

\subsection{Analytic results in the dirty limit}

I here summarize analytic results in the dirty limit $\tau\!\rightarrow\! 0$.
First, the key quantities ${\tilde K}^{0}_{N}$ are calculated by choosing
$N_{\rm cut}\!=\! 1$ in the procedure in Sec.\ IIF
of I.
The results are given by
\begin{equation}
{\tilde K}^{0}_{0}=\frac{\tilde{\varepsilon}_{n}}{\tilde{\varepsilon}_{n}^{2}+\beta^{2}} \, , 
\hspace{5mm}
{\tilde K}^{0}_{1}=\frac{\beta}{\tilde{\varepsilon}_{n}^{2}+\beta^{2}} \, . 
\end{equation}
Since $\beta^{2}$ is of the order of $1/\tau$, as shown below,
$\langle {\tilde K}^{0}_{0}\rangle$ may be approximated as
$\langle {\tilde K}^{0}_{0}\rangle\!\approx\! 
\langle 1/\tilde{\varepsilon}_{n}\!-\!\beta^{2}/\tilde{\varepsilon}_{n}^{3}
\rangle \!\approx\!
\tilde{\varepsilon}_{n}/(\tilde{\varepsilon}_{n}^{2}\!+\!
\langle \beta^{2}\rangle)$.
Using this $\langle {\tilde K}^{0}_{0}\rangle$ in Eq.\ (\ref{f_N}) and
retaining only the leading-order contributions, we obtain
\begin{equation}
\tilde{f}^{(1)}_{0}=\frac{1}{|{\varepsilon}_{n}|
+2\tau\langle\beta^{2}\rangle} \, , 
\hspace{5mm}
\tilde{f}^{(1)}_{1}=\frac{2\tau\beta\,{\rm sgn}(\varepsilon_{n})}{|{\varepsilon}_{n}|
+2\tau\langle\beta^{2}\rangle} \, . 
\label{f_Ndirty}
\end{equation}
Notice that $\tilde{f}^{(1)}_{1}$ is smaller than $\tilde{f}^{(1)}_{0}$
by $\sqrt{\tau}$. 
Substitution of Eq.\ (\ref{f_Ndirty}) into Eq.\ (\ref{Hc2})
leads to the equation for $H_{c2}$ obtained by Maki\cite{Maki64} and de Gennes:\cite{deGennes64}
\begin{equation}
\ln ({T_{c}}/{T})+\psi(1/2)-\psi\!\left(x \right) = 0\, ,
\label{Hc2dirty}
\end{equation}
where $\psi$ is the digamma function, and $x$ is defined by
\begin{equation}
x\equiv \frac{1}{2}+
\frac{\tau\langle\beta^{2}\rangle}{\pi T}= \frac{1}{2}+\frac{\tau H_{c2}}{4\pi Td}
\, ,
\end{equation}
with $d\!=\!2,3$ dimension of the system.
As shown by Maki,\cite{Maki64} Eq.\ (\ref{Hc2dirty}) can be solved near 
$T\!=\!0$ by using the asymptotic expression of $\psi(x)$ as
\begin{equation}
H_{c2}\approx \frac{d}{\tau} 
\left[1-\frac{2}{3}(\pi T)^{2}\right]\, .
\label{Hc2-dirty-0}
\end{equation}
Thus $\beta^{2}\!\propto\!H_{c2}\!\sim\!\tau^{-1}$, as assumed at the beginning.
Differentiating Eq.\ (\ref{Hc2dirty}) with respect to $T$, we obtain
\begin{equation}
\frac{dH_{c2}}{dT} = \frac{H_{c2}}{T}\!\left[1-\frac{4\pi T d}{\tau H_{c2}
\psi'(x)}\right]\, .
\end{equation}
Finally, $\kappa_{2}$ and $[\Delta_{0}(B)]^{2}$ are calculated from Eqs.\ (34b) and (36)
of I as
\begin{equation}
\kappa_{2}=\frac{d\sqrt{-\psi^{(2)}(x)}}{\sqrt{2}\tau\psi'(x)}\kappa_{0}
\hspace{2mm}
\stackrel{T\rightarrow 0}{\longrightarrow}  
\hspace{2mm} 
\frac{H_{c2}}{\sqrt{2}}\kappa_{0} 
\, ,
\label{kappa_2-dirty}
\end{equation}
\begin{equation}  
\Delta_{0}^{2} = \frac{(H_{c2}-B)\kappa_{0}^{2}}
{(2\kappa_{2}^{2}\!-\!1)\beta_{\rm A}\!+\!1}\,
\frac{4\pi Td}{\tau\psi'(x)} 
\hspace{2mm} 
\stackrel{T\rightarrow 0}{\longrightarrow} \hspace{2mm} 
\frac{(H_{c2}-B)H_{c2}\kappa_{0}^{2}}{(H_{c2}^{2}\kappa_{0}^{2}\!-\!1)\beta_{\rm A}\!+\!1} 
\, ,
\label{Delta_0-dirty}
\end{equation}
where $\kappa_{0}$ is defined by 
$\kappa_{0}\!\equiv\! \phi_{0}/2\pi\xi_{0}^{2}H_c(0)$ with $H_c(0)$ the
thermodynamic critical field at $T\!=\!0$.
Equation (\ref{kappa_2-dirty}) agrees with the result
by Caroli, Cyrot, and de Gennes.\cite{CCdG66}

Now, let us substitute Eq.\ (\ref{f_Ndirty}) into Eqs.\ (\ref{Ms3}) and (\ref{S3b})
and use Eqs.\ (\ref{Hc2-dirty-0}).
We thereby obtain
\begin{subequations}
\label{S-M-dirty}
\begin{equation}
\frac{S_{\rm s}}{S_{\rm n}}=1+\frac{dH_{c2}}{dT} \frac{3\tau\Delta_{0}^{2}}{8\pi^{3}T^{2}d}
\psi'(x)
\hspace{2mm}
\stackrel{T\rightarrow 0}{\longrightarrow} 
\hspace{2mm} 
 1-2\Delta_{0}^{2} 
\, .
\label{S-dirty}
\end{equation}
\begin{equation}
\frac{M_{{\rm sP}}}{M_{{\rm nP}}}=1+\frac{\Delta_{0}^{2}}{8\pi^{2}T^{2}}
\psi^{(2)}(x)
\hspace{2mm}
\stackrel{T\rightarrow 0}{\longrightarrow} 
\hspace{2mm} 
1-2\Delta_{0}^{2} 
\, ,
\label{Ms-dirty}
\end{equation}
\end{subequations}
Thus, ${M_{{\rm sP}}}/{M_{{\rm nP}}}$ and ${S_{{\rm s}}}/{S_{{\rm n}}}$ 
is the same at $T\!=\!0$, in agreement with Eq.\ (\ref{MS-DOS}); 
they are both determined by the zero-energy density of states.
Equation (\ref{Ms-dirty}) is the result obtained by Maki.\cite{Maki66}
Also, the expression $1\!-\!2\Delta_{0}^{2}$ for the normalized zero-energy density of states
at $T\!=\!0$
agrees with the result for the local density of states 
obtained by de Gennes.\cite{deGennes64,Fetter69}

Equation (\ref{Delta_0-dirty}) tells us that
$\Delta_{0}^{2}\!=\!(1\!-B/H_{c2})\beta_{\rm A}^{-1}$ as $T\!\rightarrow\!0$
for $\kappa_{2}\!\gg\! 1$. 
We hence find from Eqs.\ (\ref{alpha}), (\ref{S-dirty}), and (\ref{Ms-dirty})
that the initial slopes at $T\!=\!0$ for $\kappa_{2}\!\gg\! 1$ are given by
\begin{equation}
\alpha_{S}=\alpha_{\chi}=2/\beta_{\rm A}=1.72 \, .
\end{equation}
The results suggest the overall field dependence 
of $S_{\rm s}$ and $\chi_{\rm s}$ at $T\!=\!0$ which
is convex downward.
Notice that the flux-flow resistivity $\rho_{f}$ at $T\!=\!0$ also 
has the same initial slope $\alpha_{\rho}\!=\!1.72$
in the dirty limit.\cite{Kita04,Thompson70,TE70} 
These results strongly suggest that
the density of states at $\varepsilon\!=\!0$ is mainly relevant to the physical properties
of the vortex state at $T\!=\!0$.

\subsection{The case with $p$-wave impurity scattering}

If the $p$-wave impurity scattering is relevant,
the following additional terms appears 
on the right-hand side of Eq.\ (\ref{I}):
\begin{eqnarray}
d\,\frac{f \hat{\bf k} \cdot\langle \hat{\bf k}'f^{\dagger}\rangle
\!+\!\langle f\hat{\bf k}'\rangle \cdot\hat{\bf k}f^{\dagger}}{4\tau_{1}}
+d\,\frac{g\hat{\bf k}\cdot\langle \hat{\bf k}'g\rangle}{2\tau_{1}} \, ,
\end{eqnarray}
where $\langle\hat{\bf k}'g\rangle\!\equiv\!\langle\hat{\bf k}'
g(\varepsilon_{n},{\bf k}_{\rm F}',{\bf r})\rangle$, for example,
$\tau_{1}$ is the $p$-wave relaxation time,
and $\hat{\bf k}$ is the unit vector along ${\bf k}_{\rm F}$.
However, Eqs.\ (\ref{S}), (\ref{Ms}), and  (\ref{Ms1}) remain unchanged
once $I$ is modified as above.

The corresponding calculations near $H_{c2}$ may be performed 
as described in Appendix A of I.
It thereby follows that 
Eq.\ (\ref{S-M3}) and (\ref{deltaC}) are also valid
together with Eqs.\ (\ref{df_0/dB}), (\ref{f_0'}),
and (\ref{dHc2dT}),
where $\tilde{f}_{N}^{(1)}$ is now given by
\begin{eqnarray}
&&\hspace{-4mm}
\tilde{f}_{N}^{(1)}
=\frac{1}{D}
\biggl\{\!\left[1
\!-\!\frac{d}{4\tau_{1}}\langle \tilde{K}_{1}^{1}\!\sin^{2}\!\theta'\rangle\,
{\rm sgn}(\varepsilon_{n})
\right]\tilde{K}^{0}_{N}\,{\rm sgn}(\varepsilon_{n})
\nonumber \\
&& \hspace{13mm}
+\frac{d}{4\tau_{1}}{\langle \tilde{K}^{0}_{1}\!\sin\theta'\rangle\tilde{K}^{1}_{N}\sin\theta}
\biggr\}
\, ,
\label{f(1)p1}
\end{eqnarray}
with
\begin{eqnarray}
&& \hspace{-4mm}
D\equiv 
\left[1
\!-\!\frac{1}{2\tau}\langle\tilde{K}_{0}^{0}\rangle\, 
{\rm sgn}(\varepsilon_{n})\right]\!\!
\left[1
\!-\!\frac{d}{4\tau_{1}}\langle \tilde{K}_{1}^{1}\!\sin^{2}\!\theta'\rangle\,
{\rm sgn}(\varepsilon_{n})
\right]
\nonumber \\
&& \hspace{4mm}
+\frac{d}{8\tau\tau_{1}}\langle \tilde{K}_{1}^{0}\!\sin\theta'\rangle^{2} \, .
\label{f(1)pD}
\end{eqnarray}
In addition, Eq.\ (\ref{f_N'}) is to be replaced by
\begin{eqnarray}
&&\hspace{-5mm}\frac{\partial
\tilde{f}_{N}^{(1)}}{\partial \varepsilon_{n}} = -\sum_{N}\,
{\tilde K}^{N'}_{N}\tilde{f}^{(1)}_{N'}
+\frac{{\tilde K}^{0}_{N}}{2\tau}{\rm sgn}(\varepsilon_{n})
\frac{\partial\langle 
\tilde{f}_{0}^{(1)}\rangle}{\partial \varepsilon_{n}}
\nonumber \\
&&\hspace{8.5mm}
+\,d\,\frac{{\tilde K}^{1}_{N}\sin\theta}{4\tau_{1}}{\rm sgn}(\varepsilon_{n})
\frac{\partial\langle 
\tilde{f}_{1}^{(1)}\!\sin\theta'\rangle}{\partial \varepsilon_{n}} \, ,
\label{f_N'-p}
\end{eqnarray}
where
\begin{equation}
\frac{\partial\langle 
\tilde{f}_{0}^{(1)}\!\sin\theta'\rangle}{\partial \varepsilon_{n}} = -\sum_{N}\,(-1)^{N}
\langle \tilde{f}^{(1)}_{N}\tilde{\phi}^{(1)}_{N}\rangle\, {\rm sgn}(\varepsilon_{n}) \, ,
\label{f_0'sin}
\end{equation}
with
\begin{eqnarray}
&&\hspace{-4mm}
\tilde{\phi}_{N}^{(1)}
\equiv\frac{1}{D}
\biggl\{
-\!\left[1
\!-\!\frac{1}{2\tau}\langle\tilde{K}^{0}_{0}\rangle\, 
{\rm sgn}(\varepsilon_{n})\right]
\tilde{K}^{1}_{N}\sin\theta\,{\rm sgn}(\varepsilon_{n})
\nonumber \\
&& \hspace{13mm}
+\frac{1}{2\tau}\langle \tilde{K}^{0}_{1}\!\sin\theta'\rangle\tilde{K}^{0}_{N}
\biggr\}
\, .
\label{f(1)p2}
\end{eqnarray}

Finally, the analytic results in the dirty limit are the same as those given in Sec.\ IID
with a replacement of $\tau$ by the transport life time $\tau_{\rm tr}$
defined through
\begin{equation}
\frac{1}{\tau_{\rm tr}}\equiv \frac{1}{\tau}-\frac{1}{\tau_{1}}
 \, .
\label{tauTr}
\end{equation}

\subsection{\label{subsec:numproc}Numerical procedures}

I have adopted the same parameters as I and II
to express different impurity concentrations:
\begin{equation}
\xi_{\rm E}/l_{\rm tr} \equiv 1/2\pi T_{c}\tau_{\rm tr} \, , \hspace{5mm}
l_{\rm tr}/l \equiv \tau_{\rm tr}/\tau\, .
\label{parameters}
\end{equation}
Numerical calculations of Eqs.\ (\ref{S-M3}) and (\ref{deltaC})
with Eqs.\ (\ref{df_0/dB}), (\ref{f_0N'}), and (\ref{dHc2dT})
have been performed for each set of parameters by restricting every summation over 
the Matsubara frequencies for those satisfying 
$|\varepsilon_{n}|\!\leq\!\varepsilon_{c}$.
Choosing $\varepsilon_{c}\!=\!200$ is sufficient to obtain
an accuracy of $\sim\!0.1\%$ for Eqs.\ (\ref{Ms3}) and (\ref{deltaC}),
whereas $\varepsilon_{c}\!=\!20000$ ($4000$) is required for Eq.\
(\ref{S3}) in the dirty (clean) limit.
Summations over Landau levels 
have been truncated at $N\!=\!N_{\rm cut}$ where I put ${\cal R}_{N_{\rm cut}}\!=\!1$
in the calculation of ${\tilde K}^{N'}_{N}$; see Sec.\ IIF of I for the details.
Enough convergence has been obtained by choosing
$N_{\rm cut}\!=\!4$, $40$, $100$, $200$, $1500$, and $4000$ for 
$\xi_{\rm E}/l_{\rm tr}\!=\!50$, $1.0$, $0.5$, $0.1$, and $0.05$, respectively.
Finally, integrations over $\theta$ have been performed by
Simpson's formula with $N_{\rm cut}\!+\! 1$ integration
points for $0\!\leq\!\theta\!\leq\!\pi/2$.

\section{Results}
\label{sec:results}

\begin{figure}[t]
\includegraphics[width=0.8\linewidth]{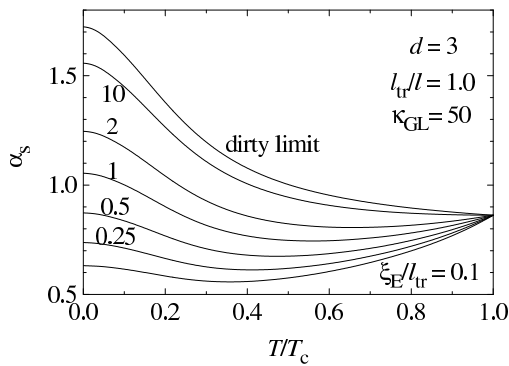}%
\caption{Slope $\alpha_{S}$ as a function of $T/T_{c}$
for different impurity concentrations
with $d\!=\!3$, $l_{\rm tr}/l\!=\!1.0$, and $\kappa_{\rm GL}\!=\!50$.
}
\label{fig:1}
\end{figure}

\begin{figure}[t]
\includegraphics[width=0.8\linewidth]{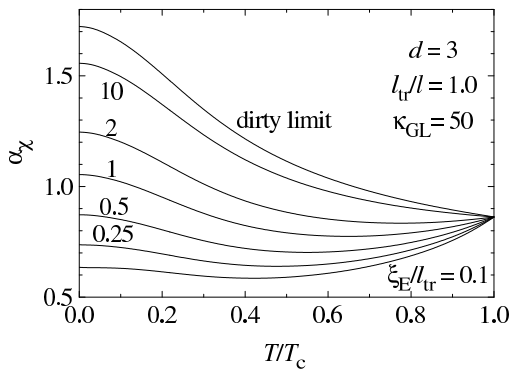}%
\caption{Slope $\alpha_{\chi}$ as a function of $T/T_{c}$
for different impurity concentrations
with $d\!=\!3$, $l_{\rm tr}/l\!=\!1.0$, and $\kappa_{\rm GL}\!=\!50$.
}
\label{fig:2}
\end{figure}

Figures \ref{fig:1} and \ref{fig:2} show temperature dependence of
$\alpha_{S}$ and $\alpha_{\chi}$ defined by Eqs.\ (\ref{alphaS}) and (\ref{alphaC}),
respectively, for different impurity concentrations parametrized by Eq.\ (\ref{parameters}).
They have been calculated in three dimensions for $l_{\rm tr}/l\!=\!1.0$ 
and $\kappa_{\rm GL}\!=\!50$.
All the curves start from the same value
$\alpha_{S}\!=\!\alpha_{\chi}\!=\!0.862$ at $T\!=\!T_{c}$
and develop differences
among different impurity concentrations at lower temperatures.
The equality $\alpha_{S}\!=\!\alpha_{\chi}$ holds at $T\!=\!0$, as shown
by Eq.\ (\ref{MS-DOS}), and the value decreases from $1.72$ in the dirty limit
to around $0.6$ for $\xi_{\rm E}/l_{\rm tr}\!=\!0.1$.
According to Eq.\ (\ref{MS-DOS}), this variation in the slope
at $T\!=\!0$ can be attributed to the dependence of the zero-energy density of states 
$N_{\rm s}(0,B)$ upon the impurity concentration. 
In particular, $N_{\rm s}(0,B)$ in the dirty (clean) limit decreases more rapidly (mildly)
than the linear behavior $N_{\rm n}(0)B/H_{c2}$ near $H_{c2}$.
From this result, we expect the overall field dependence 
of the entropy and spin susceptibility at $T\!=\!0$ which is convex
downward (upward) in the dirty (clean) limit, as realized from Eq.\ (\ref{alpha}).


The difference between $\alpha_{S}$ and $\alpha_{\chi}$ 
at finite temperatures is rather small, as expected
from $\alpha_{S}\!=\!\alpha_{\chi}$ holding at
$T\!=\!0$ and $T_{c}$.
In particular, the curves of $\alpha_{S}$ and $\alpha_{\chi}$ in the dirty limit 
depend neither on the dimensions nor $l_{\rm tr}/l$.
However, the dependence develop gradually
as the mean free path becomes longer.

\begin{figure}[t]
\includegraphics[width=0.8\linewidth]{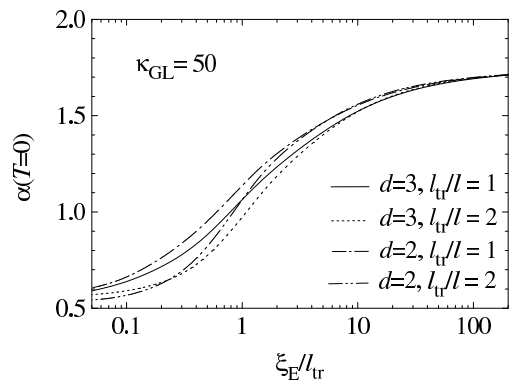}%
\caption{Slope $\alpha(T\!=\!0)\!\equiv\!\alpha_{S}(T\!=\!0)\!=\!\alpha_{\chi}(T\!=\!0)$
as a function of $\xi_{\rm E}/l_{\rm tr}$
for $d\!=\!2,3$, $l_{\rm tr}/l\!=\!1, 2$, and $\kappa_{\rm GL}\!=\!50$.
}
\label{fig:3}
\end{figure}

Figure \ref{fig:3} shows the slope $\alpha\!\equiv \!\alpha_{S}\!=\!\alpha_{\chi}$ at $T\!=\!0$
as a function of $\xi_{\rm E}/l_{\rm tr}$ for different combinations
of dimensions and impurity scatterings.
The four curves start from the same value $1.72$ in the dirty limit,
and decreases gradually through unity towards $0.5\!\sim\!0.6$ in the clean limit. 
However, we observe only small dependence of $\alpha(T\!=\!0)$ on $d$ and $l_{\rm tr}/l$.
We thus realize that the zero-energy density of states is mainly determined by the 
mean free path, and does not depend much on the dimensions nor the details of 
the impurity scattering.

It is interesting to note that the slope $\alpha_{\rho}$ 
for the flux-flow resistivity $\rho_{f}$, which was calculated previously\cite{Kita04}
for $d\!=\!2,3$, $l_{\rm tr}/l\!=\!1.0$, and $\kappa_{\rm GL}\!=\!50$,
show a complete numerical agreement at $T\!=\!0$ with the corresponding
$\alpha_{S}$ and $\alpha_{\chi}$, i.e., 
$\alpha_{S}\!=\!\alpha_{\chi}\!=\!\alpha_{\rho}$ for arbitrary
impurity concentrations at $T\!=\!0$.
This fact indicates that $\rho_{f}$ at $T\!=\!0$
is also determined by the zero-energy density of states.

\begin{figure}[b]
\includegraphics[width=0.8\linewidth]{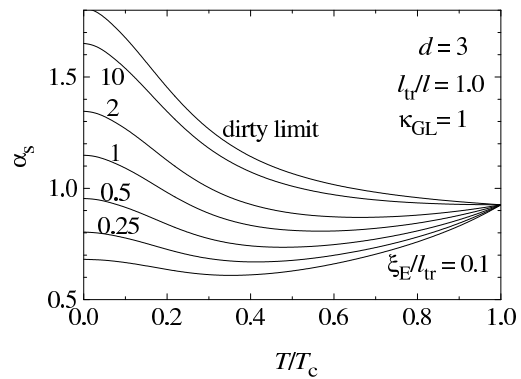}%
\caption{Slope $\alpha_{S}$ as a function of $T/T_{c}$
for different impurity concentrations
with $d\!=\!3$, $l_{\rm tr}/l\!=\!1.0$, and $\kappa_{\rm GL}\!=\!1$.
}
\label{fig:4}
\end{figure}

Next, we examine the dependence of the slopes on the Ginzburg-Landau
parameter $\kappa_{\rm GL}$.
Figure \ref{fig:4} shows the same curves as Fig.\ \ref{fig:1} near the type-I-type-II
boundary of $\kappa_{\rm GL}\!=\!1$.
Each curve is shifted upwards from the corresponding one in Fig.\ \ref{fig:1}
for $\kappa_{\rm GL}\!=\!50$,
but the quantitative difference is rather small.
This is also the case for $\alpha_{\chi}$.
Thus, the slopes $\alpha_{S}$ and $\alpha_{\chi}$
as a function of $B$ do not have large $\kappa_{\rm GL}$ dependence.

\begin{figure}[t]
\includegraphics[width=0.8\linewidth]{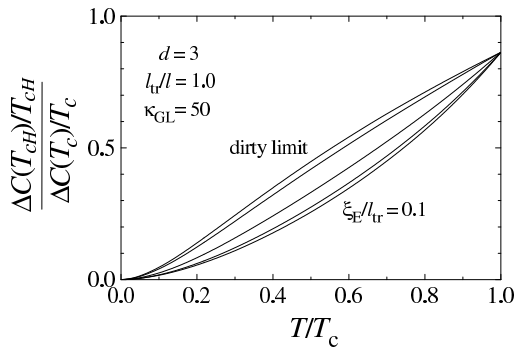}%
\caption{Specific-heat jump divided by $T$ at $H_{c2}$,
normalized by the corresponding quantity at $T\!=\!T_{c}$ and $H\!=\!0$,
as a function of $T/T_{c}$
with $d\!=\!3$, $l_{\rm tr}/l\!=\!1$, and $\kappa_{\rm GL}\!=\!50$.
The curves are for $\xi_{\rm E}/l_{\rm tr}\!=\!\infty$, $10$, $1$, $0.25$, and $0.1$
from the top to the bottom.
}
\label{fig:5}
\end{figure}

Finally, Fig.\ \ref{fig:5} plots the specific-heat jump $\Delta C$ over $T$ at $H_{c2}$
as a function of $T/T_{c}$ for different impurity concentrations
with $d\!=\!3$, $l_{\rm tr}/l\!=\!1$, and $\kappa_{\rm GL}\!=\!50$.
It is normalized by the corresponding quantity at $T\!=\!T_{c}$ and $H\!=\!0$,
i.e., $\Delta C(T_{c})/T_{c}\!=\! 1.43$ in the weak-coupling model.
The curves change gradually from almost $T$-linear overall temperature dependence 
in the dirty limit to $T^{2}$ dependence in the clean limit,
and approach zero as $\propto\! T^{2}$ at lowest temperatures.\cite{Maki65}
Although the ratio near $T_{c}$ is strongly dependent on $\kappa_{\rm GL}$ as\cite{Radebaugh66}
\begin{equation}
\lim_{T_{cH}\rightarrow T_{c}} \frac{\Delta C(T_{cH})/T_{cH}}{\Delta C(T_{c})/T_{c}}
= \frac{2\kappa_{\rm GL}^{2}}{(2\kappa_{\rm GL}^{2}-1)\beta_{\rm A}} \, ,
\end{equation}
the basic features pointed above are common among different values
of $\kappa_{\rm GL}$, $d\!=\!2,3$, and $l_{\rm tr}/l\!=1,2$.

\section{Summary}
\label{sec:summary}

The entropy and the spin susceptibility near $H_{c2}$ have
been calculated for $s$-wave type-II superconductors over all impurity concentrations. 
The results have been expressed conveniently 
using the initial slopes $\alpha_{S}$ and $\alpha_{\chi}$ defined
by Eq.\ (\ref{alpha}).
The main conclusions are summarized as follows:
(i) $\alpha_{S}\!=\!\alpha_{\chi}$ holds both at $T\!=\!0$ and $T\!=\!T_{c}$.
(ii) $\alpha_{S}\!=\!\alpha_{\chi}\!=\!0.862$ at $T\!=\!T_{c}$ for all impurity concentrations.
(iii) At $T\!=\!0$, the slope $\alpha$
decreases from $1.72$ in the dirty limit to $0.5\!\sim\!0.6$
in the clean limit.
This change is due completely to the mean-free-path dependence of 
the zero-energy density of states.
The fact also suggests variation of the overall field dependence at $T\!=\!0$
from convex downward in the dirty limit to upward in the clean limit.
(iv) The slopes have only small dependence on the dimensions and the details of the
impurity scattering.
(v) The slope $\alpha_{\rho}$ of the flux-flow resistivity $\rho_{f}$,
which was calculated 
previously,\cite{Kita04} also shows a complete numerical agreement at $T\!=\!0$
with $\alpha_{S}$ and $\alpha_{\chi}$. 
This fact indicates that the zero-energy density of states 
is also responsible for $\rho_{f}$ at $T\!=\!0$.

The $T$-linear specific-heat coefficient 
$\gamma_{s}(B)$ observed in clean 
materials\cite{Sanchez95,Nohara99,Sonier99,Hanaguri03,Hedo98,Radebaugh66,Ferreira69}
show curves of $\alpha\!<\! 1$, which are in a qualitative agreement 
with the present calculation.
On the other hand, $\gamma_{s}(B)$ for dirty samples\cite{Nohara99,Hanaguri03}
follows well-accepted linear field dependence $\propto\!B/H_{c2}$,
which is apparently in contradiction with the present result in the dirty limit.
It should be noted however that a careful experiment\cite{Gey76} on $\rho_{f}$
shows field dependence near $T\!=\!0$ which is convex downward, 
and experimentally obtained $\alpha_{\rho}$
agrees quantitatively with the dirty-limit theory.\cite{Thompson70,TE70,Kita04}
Detailed experiments on the mean-free-path dependence of $\gamma_{s}(B)$ and $\rho_{f}(B)$
are desired to remove these discrepancies.

\begin{acknowledgments}
This research is supported by a Grant-in-Aid for Scientific Research 
from the Ministry of Education, Culture, Sports, Science, and Technology
of Japan.
\end{acknowledgments}



\begin{thebibliography}{99}
\bibitem{Maki65}K. Maki, Phys. Rev. {\bf 139}, A702 (1965).

\bibitem{Maki66}K. Maki, Phys. Rev. {\bf 148}, 362 (1966).

\bibitem{CdGM64}C. Caroli, P.G. de Gennes, and J. Matricon, Phys. Lett. {\bf 9}, 
307 (1964).

\bibitem{Fetter69}A. L. Fetter and P. C. Hohenberg, in {\em Superconductivity},
edited by R. D. Parks (Dekker, NY, 1969), Vol. 2, p. 817.


\bibitem{Ichioka99}M. Ichioka, A. Hasegawa, and K. Machida, 
Phys. Rev. B {\bf59}, 184 (1999).

\bibitem{Ramirez96}A.P.Ramirez, Phys. Lett. A 211, 59 (1996).

\bibitem{Sanchez95}D. Sanchez, A. Junod, J. Muller, H. Berger, and F. Le\'vy, 
Physica B {\bf 204}, 167 (1995).

\bibitem{Nohara99}M. Nohara, M. Isshiki, F. Sakai, and H. Takagi, 
J. Phys. Soc. Jpn. {\bf 68}, 1078 (1999).

\bibitem{Sonier99}J.E. Sonier, M.F. Hundley, J.D. Thompson, and J.W.Brill,
Phys. Rev. Lett. {\bf 82}, 4914 (1999).

\bibitem{Hanaguri03}T. Hanaguri, A. Koizumi, K. Takaki, M. Nohara, H. Takagi, and K. Kitazawa,
Physica B {\bf 329}-{\bf 333}, 1355 (2003).

\bibitem{Hedo98}M. Hedo, Y. Inada, E. Yamamoto, Y. Haga,
Y. \={O}nuki, Y. Aoki, T.D. Matsuda, and H. Sato,
J. Phys. Soc. Jpn. {\bf 67}, 272 (1998).

\bibitem{Radebaugh66}R. Radebaugh and P.H. Keesom, Phys. Rev. {\bf 149}, 217 (1966).

\bibitem{Ferreira69}J. Ferreira da Silva, E.A. Burgemeister, and Z. Dokoupil, Physica
{\bf 41}, 409 (1969).

\bibitem{Kita03} T. Kita, Phys. Rev. B {\bf 68}, 184503 (2003).

\bibitem{Kita04} T. Kita, to be published in J. Phys. Soc. Jpn. {\bf 73}, No. 1 (2004); 
cond-mat/0307067.

\bibitem{Eilenberger64}G.\ Eilenberger, Z.\ Phys.\ {\bf 180} 32, (1964).

\bibitem{Lasher}G.\ Lasher, Phys.\ Rev.\ {\bf 140} A523, (1965).

\bibitem{Marcus}P.M.\ Marcus, in {\em Proceedings of the Tenth
International Conference on Low Temperature Physics}, ed.\ 
M.P.\ Malkov {\em et al.} (Viniti, Moscow, 1967) Vol.\ IIA, p.\ 345.


\bibitem{Brandt}E.H.\ Brandt, Phys.\ Stat.\ Sol.\ B {\bf 51} 345, (1972);
Phys. Rev. Lett. {\bf 78}, 2208 (1997).

\bibitem{DGR90}M.M.\ Doria, J.E.\ Gubernatis and D.\ Rainer, 
Phys.\ Rev.\ B {\bf 41}, 6335 (1990).

\bibitem{Kita98} T. Kita, J. Phys. Soc. Jpn. {\bf 67}, 2067 (1998).

\bibitem{SR83}For a review on the quasiclassical theory, 
see, e.g., J.W. Serene and D. Rainer, Phys. Rep. {\bf 101}, 221 (1983).

\bibitem{Eilenberger68} G. Eilenberger, Z. Phys. {\bf 214}, 195 (1968).

\bibitem{DGR89}M.M. Doria, J.E. Gubernatis and D. Rainer,
Phys. Rev. B {\bf 39}, 9573 (1989).

\bibitem{Yosida58}K. Yosida, Phys. Rev. {\bf 110}, 769 (1958).

\bibitem{Maki64} K. Maki, Physics {\bf 1}, 21 (1964).

\bibitem{deGennes64} P.G. de Gennes, Phys. Condens. Materie {\bf 3}, 79 (1964).

\bibitem{CCdG66} C. Caroli, M. Cyrot, and P.G. de Gennes, Solid State Commun. {\bf 4}, 17 (1966).

\bibitem{Thompson70}R. S. Thompson, Phys. Rev. B{\bf 1}, 327 (1970).

\bibitem{TE70}H. Takayama and H. Ebisawa, Prog. Theor. Phys. {\bf 44}, 1450 (1970).

\bibitem{Gey76}R. Meier-Hirmer, M. D. Maloney and W. Gey, Z. Phys. B {\bf 23} 139, (1976).

\end{thebibliography}
\end{document}